\documentclass[aps,twocolumn,prd,superscriptaddress,longbibliography,reprint]{revtex4-1}

\usepackage{amsfonts}
\usepackage{mathrsfs}
\usepackage{amsmath}
\usepackage{color}
\usepackage{graphicx}
\usepackage{bm}
\usepackage{amssymb}
\usepackage{xspace}
\usepackage{epstopdf}
\usepackage{dcolumn}
\usepackage{longtable}
\usepackage{multirow}
\usepackage{amssymb}
\usepackage{amsthm}
\usepackage[colorlinks=true, letterpaper=true, pdfstartview=FitV, linkcolor=blue, citecolor=blue, urlcolor=blue]{hyperref}

\begin{document}
\bibliographystyle{apsrev4-1}
\title{Topological hinge modes in Dirac semimetals}

\author{Xu-Tao Zeng}
\affiliation{School of Physics, Beihang University, Beijing 100191, China}

\author{Ziyu Chen}
\affiliation{School of Physics, Beihang University, Beijing 100191, China}

\author{Cong Chen}
\affiliation{School of Physics, Beihang University, Beijing 100191, China}
\affiliation{Department of Physics, The University of Hong Kong, Hong Kong, China}

\author{Bin-Bin Liu}
\affiliation{School of Physics, Beihang University, Beijing 100191, China}

\author{Xian-Lei Sheng}
\email{xlsheng@buaa.edu.cn}
\affiliation{School of Physics, Beihang University, Beijing 100191, China}

\author{Shengyuan A. Yang}
\address{Research Laboratory for Quantum Materials, Singapore University of Technology and Design, Singapore 487372, Singapore}
\address{Center for Quantum Transport and Thermal Energy Science, School of Physics and Technology, Nanjing Normal University, Nanjing 210023, China}

\begin{abstract}
Dirac semimetals (DSMs) are an important class of topological states of matter. Here, focusing on DSMs of band inversion type, we investigate their boundary modes from the effective model perspective. We show that in order to properly capture the boundary modes,  
$k$-cubic terms must be included in the effective model, which would drive an evolution of surface degeneracy manifold from a nodal line to a nodal point. Using first-principles calculations, we demonstrate that this feature and the topological hinge modes can be clearly exhibited in $\beta$-CuI. We further extend the discussion to magnetic DSMs and show that the time-reversal symmetry breaking can gap out the surface bands and hence help to expose the hinge modes in the spectrum,
which could be beneficial for the experimental detection of hinge modes.
\end{abstract}

\maketitle

\section{Introduction}

The study of topological states and topological materials is an important research topic in the past two decades~\cite{hasan2010,qi2011,shen2012,bernevig2013,bansil2016,chiu2016,yang2016,dai2016a,burkov2016,armitage2018}.
An important property of topological states is the bulk-boundary correspondence, i.e., the nontrivial topology in the bulk of a system would manifest as protected modes at the boundary.
For example, a two-dimensional (2D) quantum anomalous Hall insulator features chiral zero-modes at its 1D edges~\cite{haldane1988}. As another example, 3D Weyl semimetals have protected surface Fermi arcs connecting the protections of bulk Weyl points with opposite chirality~\cite{wan2011,armitage2018}.
The existence of surface Fermi arcs can be argued by considering a cylindrical surface in the Brillouin zone (BZ) that encloses one Weyl point~\cite{wan2011}.
By the Gauss Law, this 2D sub-system is essentially a 2D quantum anomalous Hall insulator, and the corresponding chiral zero-mode traces out a Fermi arc on the surface when we vary the radius of the cylinder.

Dirac semimetals (DSMs) are an important class of topological states that are closely related to Weyl semimetals~\cite{young2012,wang2012a,wang2013}.
In a DSM, the bands cross at isolated Dirac points at Fermi level.
Each Dirac point is fourfold degenerate and can be regarded as formed by merging together a pair of Weyl points with opposite chirality.
Because of this, a Dirac point does not have a net chirality (or a nontrivial Chern number).
Previous works have shown that there are two types of Dirac points according to their formation mechanism~\cite{young2012,armitage2018}.
One type is the essential Dirac points, whose existence is enforced by certain nonsymmorphic space group symmetry~\cite{young2012,steinberg2014}.
The other type is the accidental Dirac points, which is associated with band inversion in a region of the BZ~\cite{wang2012a,wang2013}. On the experimental side, the latter type attracted more interest, because it finds good material realizations, such as Na$_3$Bi and Cd$_3$As$_2$, and also because it hosts interesting boundary modes~\cite{liu2014b,liu2014,neupane2014,jeon2014,borisenko2014,liang2015,xu2015a,xiong2015a}.
Initial first-principles calculations showed that Na$_3$Bi and Cd$_3$As$_2$ have surface Fermi arcs connecting the projections of bulk Dirac points~\cite{wang2012a,wang2013}, similar to those in Weyl semimetals.
However, subsequent studies pointed out that such surface arcs are not protected~\cite{kargarian2016}.
More recently, with the development of the concept of higher-order topology\cite{zhang2013,benalcazar2017,benalcazar2017a,langbehn2017,song2017,schindler2018a,schindler2018,sheng2019,wieder2020,wang2020a,ghorashi2020,chen2022}, Wieder \emph{et al.} found that these DSMs actually have a second-order topology with hinge Fermi arcs~\cite{wieder2020}.

In this work, we focus on this type of DSMs with band inversions and investigate the evolution of boundary modes from low-energy effective models.
We show that in order to correctly capture the topology and boundary modes, the effective model must include terms beyond the second order in the momentum.
Particularly, with the inclusion of $k$-cubic terms, there is an evolution of the surface degeneracy manifold from an open nodal line to a nodal point.
This understanding offers guidance to search for materials with hinge modes that can be more readily probed in practice.
We show that this is the case for $\beta$-CuI.
Its hinge modes are directly exposed in first-principles calculations.
We further extend the discussion to magnetic DSMs and show that the time reversal symmetry breaking can completely gap out the surface bands while maintaining the hinge modes, which could be beneficial for the experimental detection of hinge states.
Since effective models are widely used for understanding topological states, our findings have important implications on theoretical studies based on the such models.
The results also point to concrete materials for which the topological hinge modes can be verified in experiment.

\section{Effective model analysis}\label{sec:model}

DSMs with band inversions such as Na$_3$Bi and Cd$_3$As$_2$ share similar low-energy band structures~\cite{wang2012a,wang2013}.
They feature band inversion around a high-symmetry point (such as $\Gamma$) in the BZ, and a pair of Dirac points are protected on a rotational axis that passes through the high-symmetry point.
The commonly used low-energy effective model to study these DSMs is
\begin{equation}
  \label{EQ:DSM_k2}
  H_0(\bm k)=\varepsilon(\bm{k})+M(\bm{k})\sigma_zs_0+Ak_x\sigma_xs_z-Ak_y\sigma_ys_0,
\end{equation}
where the momentum $\bm k$ is measured from the band inversion high-symmetry point, $\sigma_i$ and $s_i$ are two sets of Pauli matrices, the functions $\varepsilon({\bm k})=C_{0}+C_{1}k_{z}^{2}+C_{2}(k_x^{2}+k_y^2)$, $M({\bm k})=M_{0}-M_{1}k_{z}^{2}-M_{2}(k_x^{2}+k_y^2)$, and $C$'s, $M$'s, and $A$ are real model parameters.
This model is expanded to the $k$-square order, which can describe the band inversion feature at the $k=0$ point if we require the $M$'s share the same sign.
Without loss of generality, we assume $M_0, M_1, M_2>0$.

As we shall show in a while, the conventional model in (\ref{EQ:DSM_k2}) is not sufficient to capture the second-order topology and the correct boundary modes.
To remedy this, expansion beyond the $k$ square order is needed.
Here, we shall include the $k$ cubic terms, which are sufficient for the task.

Obviously, the form of the $k$ cubic terms depends on the crystal symmetry of the material to be considered.
To be specific, let's consider the constraint of $D_{6h}$ point group symmetry, which applies to the material Na$_3$Bi.
In Appendix~\ref{appendix:AppC4}, we also present the analysis for the $D_{4h}$ point group (applying to Cd$_3$As$_2$), which leads to slightly different terms, but the qualitative results regarding their influence on the topology are not affected. Considering the constraint from time-reversal symmetry $\mathcal{T} = -i\sigma_0s_y\mathcal{K}$ ($\mathcal{K}$ the complex conjugation) and the generators of the $D_{6h}$ group: $\mathcal{P} = \sigma_zs_0$, $\mathcal{M}_x = i\sigma_0s_x$ and $C_{6z} = e^{i(\pi/3)\hat{J_z}/\hbar} = e^{i(\pi/3)(2\sigma_0-\sigma_z)s_z}$,
the symmetry-allowed $k$ cubic terms include
\begin{equation}
  H_1(\bm k)=B k_z[(k_x^2-k_y^2)\sigma_xs_x+2k_xk_y\sigma_xs_y],
  \label{EQ:DSM_k3}
\end{equation}
with $B$ a real parameter. Note that besides $H_1$, there are additional $k$-cubic terms proportional to the last two terms in (\ref{EQ:DSM_k2}) timed by $k_i^2$. However, these terms are not important for our discussion, so they are neglected here.

The spectrum of the effective model $H=H_0+H_1$ can be readily solved, which is given by
\begin{equation}\label{Ek}
  E_\pm({\bm k})= \varepsilon({\bm k})\pm\sqrt{M({\bm k})^{2}+A^{2}k_+k_-+|B k_zk_{+}^2|^2},
\end{equation}
where $k_{\pm}=k_{x}\pm ik_{y}$, and each band is doubly degenerate due to the $\mathcal{PT}$ symmetry. The bands cross at two Dirac points
located at $(0,0,\pm k_D)$ on the high-symmetry axis, with $k_D=\sqrt{M_0/M_1}$. Around each Dirac point, the band dispersion is linear in $k$ at the leading order. For example, expanding the dispersion at $(0,0,+k_D)$, we have $E(\bm q)=\pm\sqrt{A(q_x^2+q_y^2)+4M_0M_1 q_z^2}\sim q$, where $\bm q$ and the energy are measured from the Dirac point.

\begin{figure}[tb]
  \includegraphics[width=8cm]{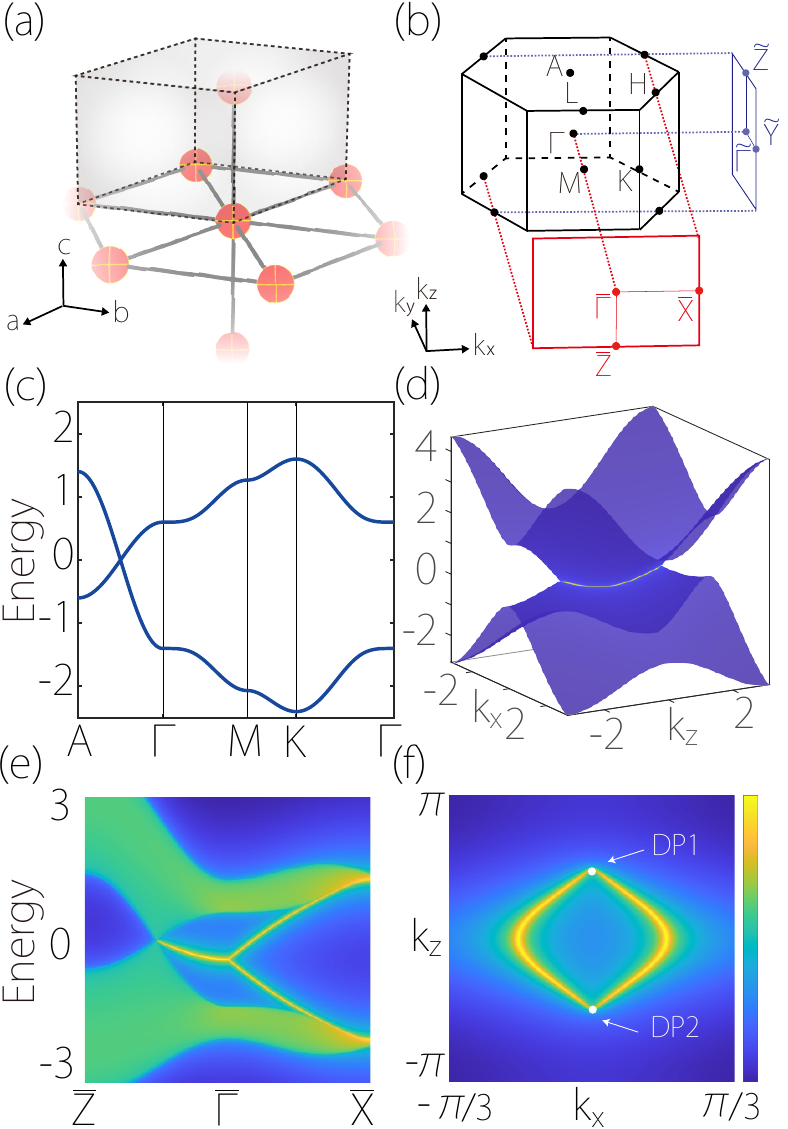}
  \caption{ DSM effective model without the $k$-cubic terms [Eq.~(\ref{EQ:DSM_k2})]. (a) We discretize the model on a 3D hexagonal lattice. (b) The corresponding BZ. (c) Bulk band structure. Here, each band is twofold degenerate.   (d) Surface band dispersion on a side surface. There is a surface nodal line form by the crossing of surface bands, which connects the projections of two bulk Dirac points.  (e) The corresponding surface spectrum along high symmetry path and (f) the constant energy slice at Fermi level. Here, we take the model parameters as $C_0 = 1, C_1 = 0.2, C_2 = 0, M_0 = 1, M_1 = 0.5, M_2 = 0.5, \text{and}\ A = 1$.}
  \label{fig:DSM_k2_surf}
\end{figure}
Now, we analyze the boundary modes of this effective model. First, it is noted that the $k$-cubic terms in $H_1$ do not affect the bulk Dirac point features.
For instance, if we put $B=0$ in (\ref{Ek}), one finds that the location of the Dirac points and the leading order dispersion are not affected at all, which seems to imply that $H_1$ is inessential.
Hence, let's first consider the surface spectrum by neglecting the $H_1$ term.
{The calculation results are presented in Fig.~\ref{fig:DSM_k2_surf}. Here, to study a surface, we discretize the model on a hexagonal lattice
as in Fig.~\ref{fig:DSM_k2_surf}(a).
The bulk band structure in Fig.~\ref{fig:DSM_k2_surf}(c) captures the low-energy features, particularly the Dirac points on the $\Gamma$-$A$ path.}
In Fig.~\ref{fig:DSM_k2_surf}(d, e), we plot the calculated spectrum for the side surface normal to $\hat{y}$, where the projections of the two bulk Dirac points can be well distinguished.
In Fig.~\ref{fig:DSM_k2_surf}(f), one clearly observes a pair of surface Fermi arcs connecting the two projected Dirac points, which are similar to the previous first-principles results on Na$_3$Bi and Cd$_3$As$_2$~\cite{wang2012a,wang2013}.
These Fermi arcs are formed by the cutting of Fermi energy with the surface bands indicated in Fig.~\ref{fig:DSM_k2_surf}(e). One can see that the surface bands linearly cross on the $\bar{\Gamma}$-$\bar{Z}$ path in the surface BZ between the surface protections of Dirac points at $(0,\pm k_D)$, which form a nodal line connecting the projected Dirac points in the surface band structure. This picture can be better visualized in Fig.~\ref{fig:DSM_k2_surf}(d), which maps out the surface band dispersion.

The surface spectrum for $H_0$ can be understood in the following. Consider a slice in the BZ with constant $k_z=\lambda$ for $H_0$, which constitutes a 2D sub-system $\tilde{H}_0^\lambda(k_x,k_y)$ labeled by $\lambda$. We have
\begin{equation}
  \tilde{H}_0^\lambda=\varepsilon(k_x,k_y,\lambda)+M(k_x,k_y,\lambda)\sigma_zs_0+Ak_x\sigma_xs_z-Ak_y\sigma_ys_0.
\end{equation}
One finds that this 3D model has exactly the same form as the famous Bernevig-Hughes-Zhang model~\cite{bernevig2006a} for 2D topological insulators.
Particularly, the model is topologically nontrivial for $|\lambda|<k_D$, i.e., for a 2D slice in the region between the two Dirac points, which is consistent with the assumed band inversion feature around $\bm k=0$.
Thus, each constant $k_z$ slice between the two Dirac points is effectively a 2D topological insulator, which has a pair of 1D edge bands forming a Dirac type crossing.
The crossing traces out the surface nodal line connecting the two Dirac points on a side surface.
This clarifies the origin of the surface spectrum of $H_0$ in Fig.~\ref{fig:DSM_k2_surf}(d-f).

It must be noted that a conventional topological insulator requires the protection of the time reversal symmetry.
In the 2D model $\tilde{H}_0^\lambda$, we have an anti-unitary symmetry $\mathcal{T}^*=-i\sigma_0s_y\mathcal{K}$, which resembles but is not the true time reversal symmetry for $k_z\neq 0$, because in the 3D system, time reversal operation should also reverse the sign of $k_z$.
It follows that the surface spectrum in Fig.~\ref{fig:DSM_k2_surf}(d) is enabled by an \emph{emergent} symmetry ($\mathcal{T}^*$) limited to $H_0$, but not protected by any true symmetry of the original system.
As a result, the surface bands with a nodal line represents a critical state susceptible to perturbations from higher-order terms.

\begin{figure}[tb]
  \includegraphics[width=8cm]{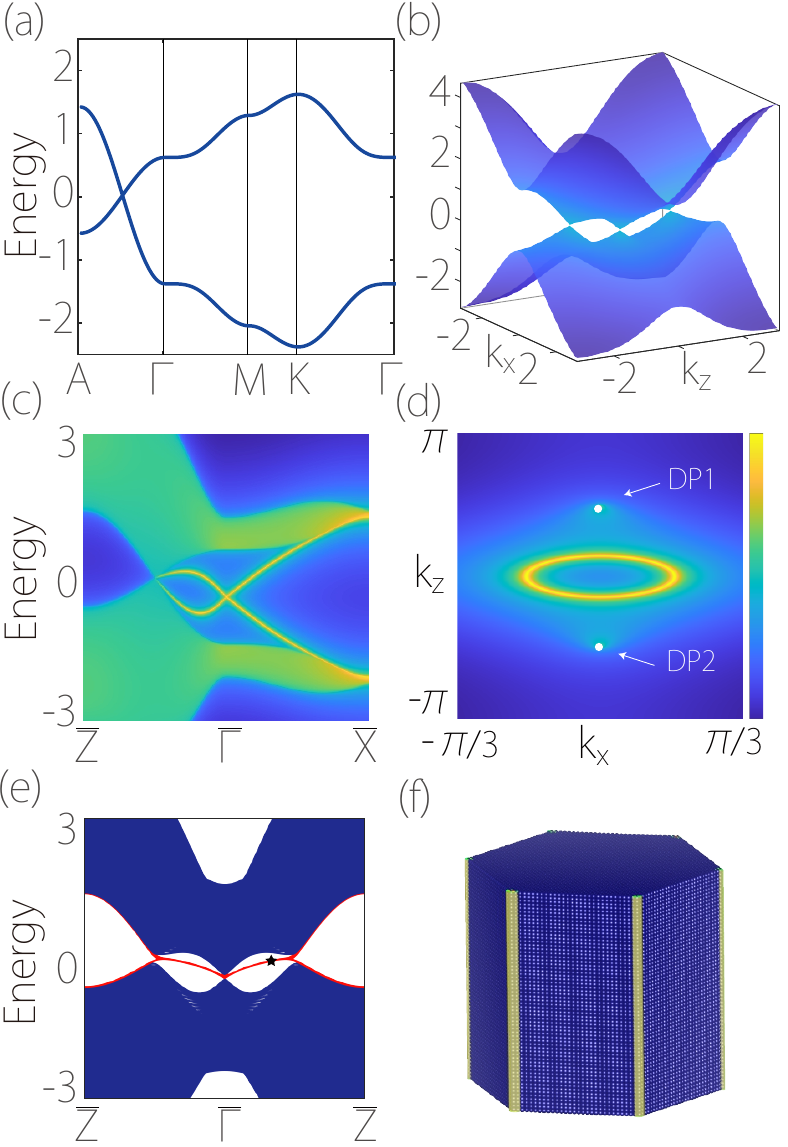}
  \caption{ DSM effective model with cubic terms included [Eq.~(\ref{EQ:DSM_k3}) $+$ Eq.~(\ref{EQ:DSM_k2})]. (a) Bulk band structure. (b) Surface band dispersion on a side surface. There is a Dirac cone at the surface BZ center. (c) Surface spectrum and (d) constant energy slice at Fermi level for the side surface. (e) Spectrum of a 1D hexagonal tube geometry (with 30 cell length of an edge) as shown in (f). The hinge modes are indicated by the red lines. (f) Spatial distribution of the hinge mode marked by the star in (e). Here, we take the parameters as $C_0 = 1, C_1 = 0.2, C_2 = 0, M_0 = 1, M_1 = 0.5, M_2 = 0.5, \text{and}\ A = 1$.}
  \label{fig:DSM_k3_surf}
\end{figure}

Next, we show that restoring the $k$-cubic terms in $H_1$ helps to capture the correct topology.
Note that by putting $k_z=\lambda$ in $H_1$, we obtain its contribution to the 2D sub-system of a constant $k_z$ slice:
\begin{equation}\label{eq5}
  \tilde{H}_1^\lambda=B\lambda[(k_x^2-k_y^2)\sigma_xs_x+2k_xk_y\sigma_xs_y].
\end{equation}
Clearly, $\tilde{H}_1^\lambda$ breaks the emergent symmetry $\mathcal{T}^*$ of $\tilde{H}_0^\lambda$.
In other words, if we treat $\tilde{H}_0^\lambda$ as describing a 2D $\mathcal{T}^*$-invariant topological insulator, $\tilde{H}_1^\lambda$ can be regarded as perturbations that break the effective time reversal symmetry.
Consequently, the 1D Dirac type crossing in the edge bands for $\tilde{H}_0^\lambda$ would open a gap.
This is confirmed by the calculated surface spectrum in Fig.~\ref{fig:DSM_k3_surf}(b) by including the $H_1$ term, which destroys the surface nodal line.
It should be noted that the $k_z=0$ slice is special as it preserves the true time reversal symmetry, so it remains a 2D topological insulator with gapless edge bands.
For the 3D system, this means that although the surface nodal line is destroyed, there is still a robust nodal point of the surface bands at $\bar{\Gamma}$.

This feature can also be understood from another perspective.
Note that the bulk Dirac points are protected by the rotational symmetry on the $k_z$ axis.
They can be gapped out by breaking the rotational symmetry while preserving $\mathcal{T}$.
Then the system would transform to a 3D strong topological insulator because of the assumed band inversion at $\Gamma$.
It is well known that a 3D topological insulator features Dirac-cone type surface bands.
This explains the Dirac type surface dispersion in Figs.~\ref{fig:DSM_k2_surf}(b-d), and the nodal point is just the neck point of the surface Dirac cone.
This discussion clarifies the important role played by $H_1$, under which the surface bands evolve from Fig.~\ref{fig:DSM_k2_surf}(d) with a nodal line to Fig.~\ref{fig:DSM_k3_surf}(b) with a Dirac cone.
Inspecting the Fermi contour at the surface, the Fermi arcs in Fig.~\ref{fig:DSM_k2_surf}(f) would generally transform into a closed loop in Fig.~\ref{fig:DSM_k3_surf}(d), similar to that in a 3D strong topological insulator.

\begin{figure}[tb]
  \includegraphics[width=8cm]{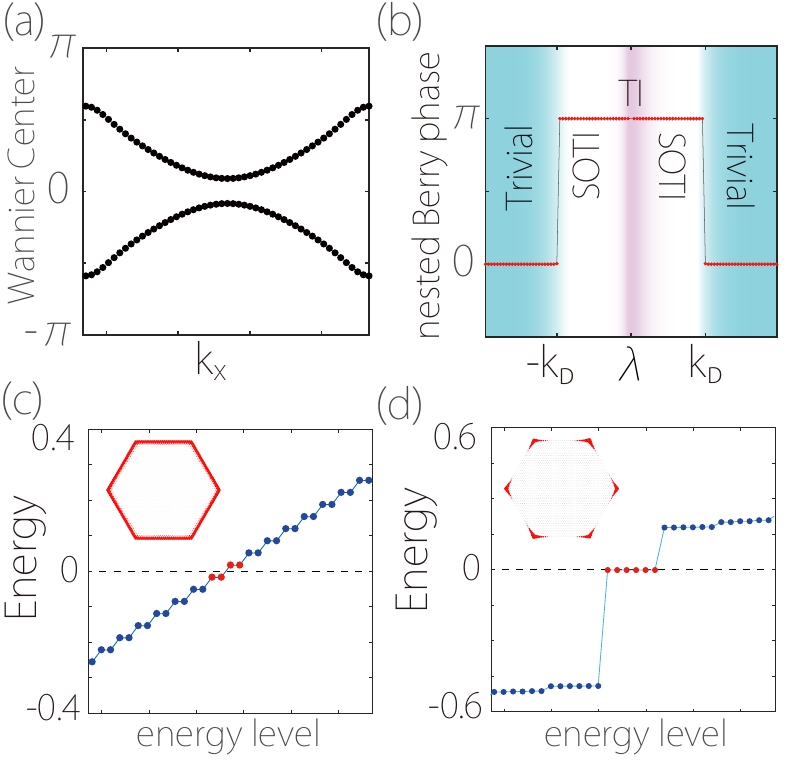}
  \caption{Results for the effective 2D Hamiltonian $\tilde{H}^\lambda(k_x,k_y)=\tilde{H}_0^\lambda+\tilde{H}_1^\lambda$ when $\lambda = \pi/3$. (a) Evolution of the Wannier centers for the occupied bands. (b) Nested Berry phase when $\lambda$ varies along $k_z$. The system has a nontrivial second-order topology for $\lambda$ between the two Dirac points. (c, d) Spectra for the nanodisk geometry (c) without and (d) with $\tilde{H}_1^\lambda$. The insets show the distribution of the states marked in red in the spectra. Here, we take the parameters as $C_0 = 1, C_1 = 0.2, C_2 = 0, M_0 = 1, M_1 = 0.5, M_2 = 0.5, \text{and}\ A = 1$.}
  \label{fig:DSM_2D}
\end{figure}

We have shown that by including the $k$-cubic terms, the 2D sub-system described by $\tilde{H}^\lambda(k_x,k_y)=\tilde{H}_0^\lambda+\tilde{H}_1^\lambda$ with $|\lambda|<k_D$ and $\lambda\neq 0$ is no longer a 2D conventional topological insulator. Both its bulk and its edge are gapped. Nevertheless, the band inversion feature is still maintained in the model, and we will show that $\tilde{H}^\lambda$ corresponds to a 2D second-order topological insulator. The second-order topology can be inferred from the nested Wilson loop calculation~\cite{benalcazar2017}. In Fig.~\ref{fig:DSM_2D}(b), we plot the obtained nested Berry phase as a function of $\lambda$. One observes that the result is nontrivial (trivial) for $|\lambda|<k_D$ ($>k_D$). Thus, each constant $k_z(\neq 0)$ slice of the BZ between the two Dirac points is effectively a 2D second-order topological insulator.

A 2D second-order topological insulator should have protected corner modes. We implement $\tilde{H}^\lambda$ on a hexagonal lattice and plot the calculated spectrum for a nanodisk geometry in Fig~\ref{fig:DSM_2D}(c, d). Here, we take $\lambda= \pi/3$. When we put $B=0$, i.e., drop the $k$-cubic terms, the zero-modes are distributed throughout the edge of the disk [Fig.~\ref{fig:DSM_2D}(c)]. This is the critical state, for which the system resembles the conventional topological insulator with gapless edge modes. As soon as we turn on the $k$-cubic terms, the edge becomes gapped and the zero-modes are localized at the corners of the disk [Fig.~\ref{fig:DSM_2D}(d)], confirming the second-order topology.

Since $\tilde{H}^\lambda$ is a constant $k_z$ slice of the DSM, its corner modes would constitute the hinge modes at hinges between the side surfaces of a 3D DSM. To explicitly demonstrate this, we consider a tube geometry as shown in Fig.~\ref{fig:DSM_k3_surf}(f). The obtained spectrum in plotted in Fig.~\ref{fig:DSM_k3_surf}(e), in which the hinge modes are marked with red color. In Fig.~\ref{fig:DSM_k3_surf}(f), we verify that these modes are indeed distributed at the hinges between the side surfaces of the system.

From the model study, we have seen that the $k$-cubic terms are indispensable for describing the correct boundary modes of the DSM. On the 2D surface, the generic Fermi contour is a Fermi loop from the Dirac-cone surface bands. The bulk band inversion leads to second-order topology with hinge modes bounded by the projected Dirac points on the 1D hinges between side surfaces.

\section{Material example}

The analysis in the last section shows that to better visualize the hinge modes, the system needs to have sizable $k$-cubic terms. In materials Na$_3$Bi and Cd$_3$As$_2$, the cubic terms are relatively small, which makes the surface Fermi contour still close to Fermi arcs. And the hinge modes there coexist in energy with the surface modes for a fixed $k_z$, making it difficult to resolve the hinge modes in the spectrum.

Here, we show that $\beta$-CuI is a good candidate to probe the hinge modes. The previous work by Le \emph{et al.}~\cite{le2018} has revealed
$\beta$-CuI as a DSM formed by band inversion. Here, we find that this material has sizable $k$-cubic terms, and we shall directly investigate its hinge modes.

\begin{figure}[tb]
  \includegraphics[width=8cm]{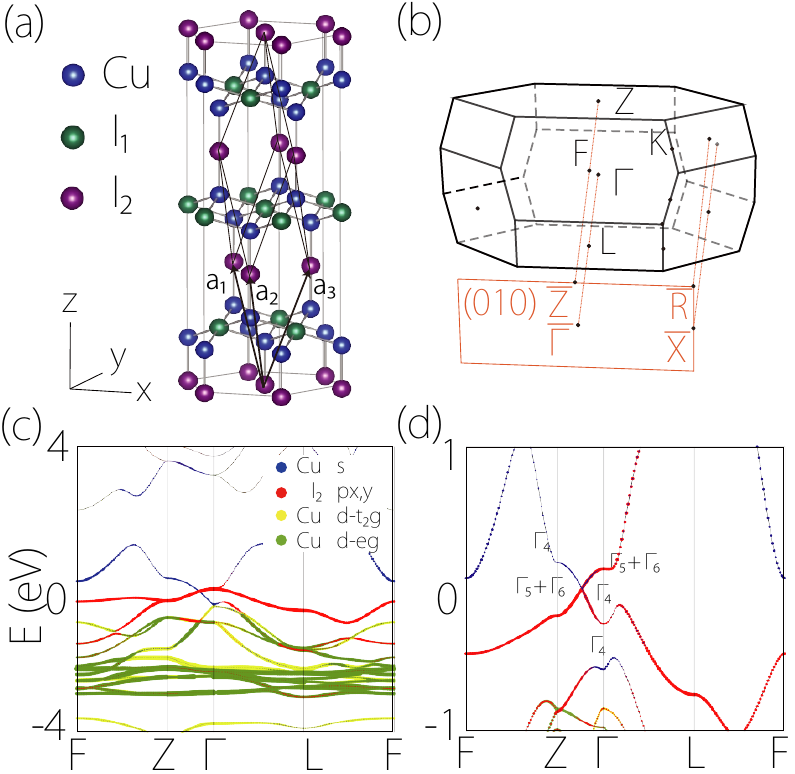}
  \caption{ (a) Crystal structure of hexagonal $\beta$-CuI. (b) The first BZ of $\beta$-CuI and its projected surface BZ on (010) planes.
    (c, d) Band structure of $\beta$-CuI (c) without and (d) with spin-orbit coupling. }
  \label{fig:structure}
\end{figure}

As illustrated in Fig.~\ref{fig:structure}(a), the structure of $\beta$-CuI belongs to the space group $R\bar{3}m$ (No.~166), same as the famous topological insulator Bi$_2$Se$_3$ family. From the crystal field environment, one observes that the iodine atoms can be classified as two types denoted as I$_1$ and I$_2$, where I$_1$ is octahedrally coordinated by six Cu atoms forming a sandwich ABC tri-layer stacking, while I$_2$ connects two Cu atoms parallel to the $c$ axis separating the Cu-I$_1$-Cu sandwich layer. The relaxed lattice constants are $a = 4.3710 $ \AA~and $c = 20.8611 $ \AA~in the hexagonal lattice description (see Appendix~\ref{appendix:method} for the computation approach), which are in good agreement with the experimental results ($a = 4.2986$\AA, $c = 21.4712$ \AA)~\cite{shan2009}. The Wyckoff positions of Cu, I$_1$ and I$_2$ are $6c$ (0, 0, 0.1246), $3a$ (0, 0, 0) and $3b$ (0, 0, 0.5), respectively.

In $\beta$-CuI, the $p$ orbitals of I$_1$ atoms and the $p_z$ orbitals of I$_2$ atoms are strongly affected by the crystal fields from the surrounding Cu atoms and are repelled away from the Fermi level. Meanwhile, due to the positive valence of Cu, the $d$ orbitals of Cu are completely filled and are located at around $-2.5$ eV. Therefore, near the Fermi level, the valence and conduction bands are mainly contributed by the I$_2$-5$p_{x,y}$ and Cu-4$s$ orbitals. Our first-principles result confirms this analysis. Figure~\ref{fig:structure}(c) shows the band structure and projected density of states (PDOS) of $\beta$-CuI without spin-orbit coupling (SOC). Around the Fermi energy, there is an energy band inversion at the $\Gamma$ point, caused by the Cu-4$s$ and the I$_2$-5$p_{x,y}$ orbitals. The Cu-4$s$ bands are about 0.47 eV lower than the I$_2$-5$p_{x,y}$ bands, and there is a band crossing point along the $\Gamma$-$Z$ line.
After turning on SOC, the band inversion at $\Gamma$ is enhanced to 0.77 eV, and the band crossing along $\Gamma$-$Z$ still exists [Fig.~\ref{fig:structure}(d)]. Each band here is doubly degenerate due to $\mathcal{PT}$. The irreducible representations of the two crossing bands belong to $\Gamma_{4}$ and $\Gamma_5 \oplus \Gamma_6$ of $C_{3v}$ group along $\Gamma$-$Z$, respectively. Therefore, the crossing point is a fourfold Dirac point, consistent with the previous result~\cite{le2018}.



\begin{figure}[htb]
  \includegraphics[width=8cm]{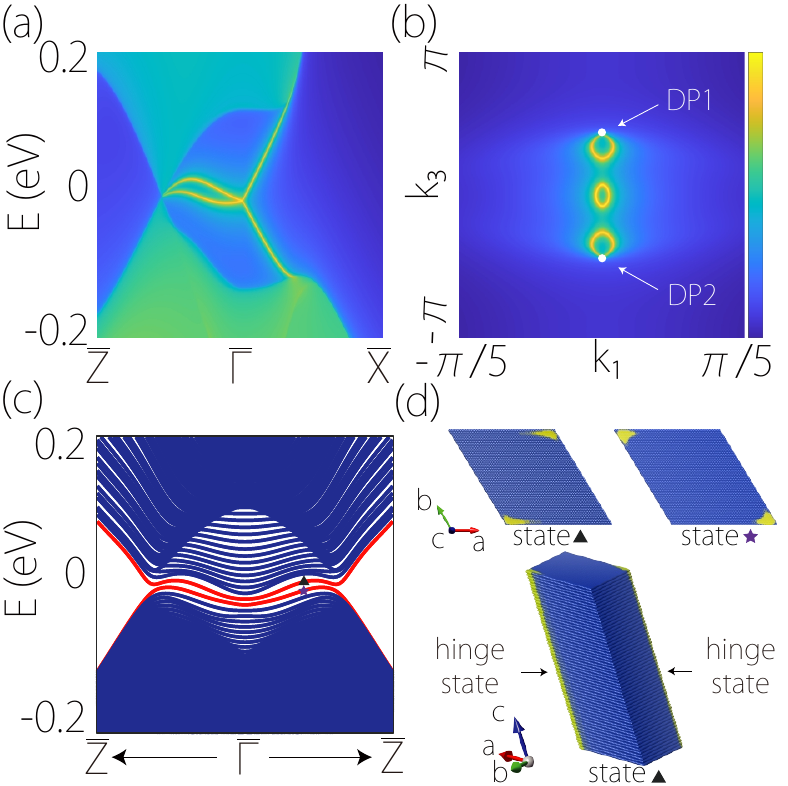}
  \caption{ (a, b) Projected spectrum and the Fermi contours for the (010) surface. (c) Spectrum for the 1D tube geometry of $\beta$-CuI as shown in (d). Here, each side of the tube cross section has a length of 60 unit cells. The hinge modes are highlighted by the red lines.
    (d) Spatial distribution of two hinge modes marked in (c). }
  \label{fig:CuI_surf}
\end{figure}

Now, we turn to the surface spectrum of $\beta$-CuI. Figure~\ref{fig:CuI_surf}(a) shows the calculated surface spectrum for the (100) surface. One observes features similar to those in Fig.~\ref{fig:DSM_k3_surf}(c).
Particularly, one can see the large splitting of the nodal line on the $\bar{\Gamma}$-$\bar{Z}$ path between the projected Dirac points, and the Fermi contour takes the form of a loop rather than arcs [Fig.~\ref{fig:CuI_surf}(b)]. These evidences indicate sizable $k$-cubic terms which break the effective $\mathcal{T}^*$ symmetry.


The surface spectrum in Fig.~\ref{fig:CuI_surf}(a) suggests that there is a good chance to resolve the hinge modes in the system. To calculate the hinge spectrum, we consider the tube geometry shown in Fig.~\ref{fig:CuI_surf}(d). The result is plotted in Fig.~\ref{fig:CuI_surf}(c). Indeed, we find two hinge bands within the surface band gap bounded by the projected Dirac points. By checking the wave function distribution, we verify that these modes are located at the hinges of the sample, as shown in Fig.~\ref{fig:CuI_surf}(d). These hinge modes manifest the second-order topological character of $\beta$-CuI.


Finally, let's construct the $k\cdot p$ effective model for $\beta$-CuI. $\beta$-CuI has the $D_{3d}$ point group symmetry.
The symmetry-constrained model is slightly more complicated than that discussed in the last section, but the qualitative features are the same.
Choosing the basis at $\Gamma$ as $|S_{1/2}^+, \pm1/2 \rangle$ and $| P_{3/2}^-, \pm3/2 \rangle$,
the symmetry generators are represented as $\mathcal{P} = \sigma_zs_0$, $\mathcal{M}_x = i\sigma_0s_x$, $C_{3z} = e^{i(2\pi/3)\hat{J_z}/\hbar} = e^{i(2\pi/3)(2\sigma_0-\sigma_z)s_z}$ and $\mathcal{T} = -i\sigma_0s_y\mathcal{K}$.

Then, the symmetry allowed effective model can be obtained as
\begin{equation}
  H(\bm{k})=H_0+H_1,\label{Eq:CuIDSM1}
\end{equation}
where $H_0$ contains terms up to $k$-square order
\begin{equation}
  \begin{split}
    H_0&=\varepsilon(\bm{k})+M(\bm{k})\sigma_zs_0+A_0\left(k_x\sigma_xs_z-k_y\sigma_ys_0\right)\\&\ +D_0\left(k_x\sigma_xs_x-k_y\sigma_xs_y\right),\\
  \end{split}
\end{equation}
and $H_1$ contains $k$-cubic terms
\begin{equation}
  \begin{split}
    H_1&=B_1 k_z[(k_x^2-k_y^2)\sigma_xs_x+2k_xk_y\sigma_xs_y]\\&\ +B_2 k_z[(k_x^2-k_y^2)\sigma_xs_z+2k_xk_y\sigma_ys_0]\\
    &\ +[A_1 k_z^2+A_2(k_x^2+k_y^2)](k_x\sigma_xs_z-k_y\sigma_y s_0)\\
    &\ +[D_{1}k_{z}^{2}+D_{2}(k_x^{2}+k_y^2)](k_x\sigma_xs_x-k_y\sigma_xs_y).
  \end{split}
\end{equation}
Here, $\varepsilon(\bm{k})$ and $M(\bm{k})$ have the same expression as in Eq.~(\ref{EQ:DSM_k2}). The model parameters can be obtained from fitting the first-principles band structure in Fig.~\ref{fig:structure}(d). We obtain that
$C_{0}=-0.0518\ \mathrm{eV}$,
$C_{1}=0.6661\ \mathrm{eV}\cdot\text{\AA}^{2}$,
$C_{2}=3.1243\ \mathrm{eV} \cdot \text{\AA}^{2}$,
$M_{0}=0.1930\ \mathrm{eV}$,
$M_{1}=4.9640\ \mathrm{eV}\cdot\text{\AA}^{2}$,
$M_{2}=0.8866\ \mathrm{eV}\cdot\text{\AA}^{2}$,
$A_{0}=-1.5556\ \mathrm{eV}\cdot\text{\AA}$,
$A_{1}=-0.0937\ \mathrm{eV}\cdot\text{\AA}^{3}$,
$A_{2}=-0.8030\ \mathrm{eV}\cdot\text{\AA}^{3}$,
$D_{0}=0.2264\ \mathrm{eV}\cdot\text{\AA}$,
$D_{1}=-0.0570\ \mathrm{eV}\cdot\text{\AA}^{3}$,
$D_{2}=8.9368\ \mathrm{eV}\cdot\text{\AA}^{3}$,
$B_1 =6.7806\ \mathrm{eV}\cdot\text{\AA}^{3}$, and
$B_2 =1.4844\ \mathrm{eV}\cdot\text{\AA}^{3}$.
The result shows that the $k$-cubic terms are sizable for $\beta$-CuI.

\section{Magnetic Dirac semimetal}

\begin{figure}[bt]
  \includegraphics[width=8cm]{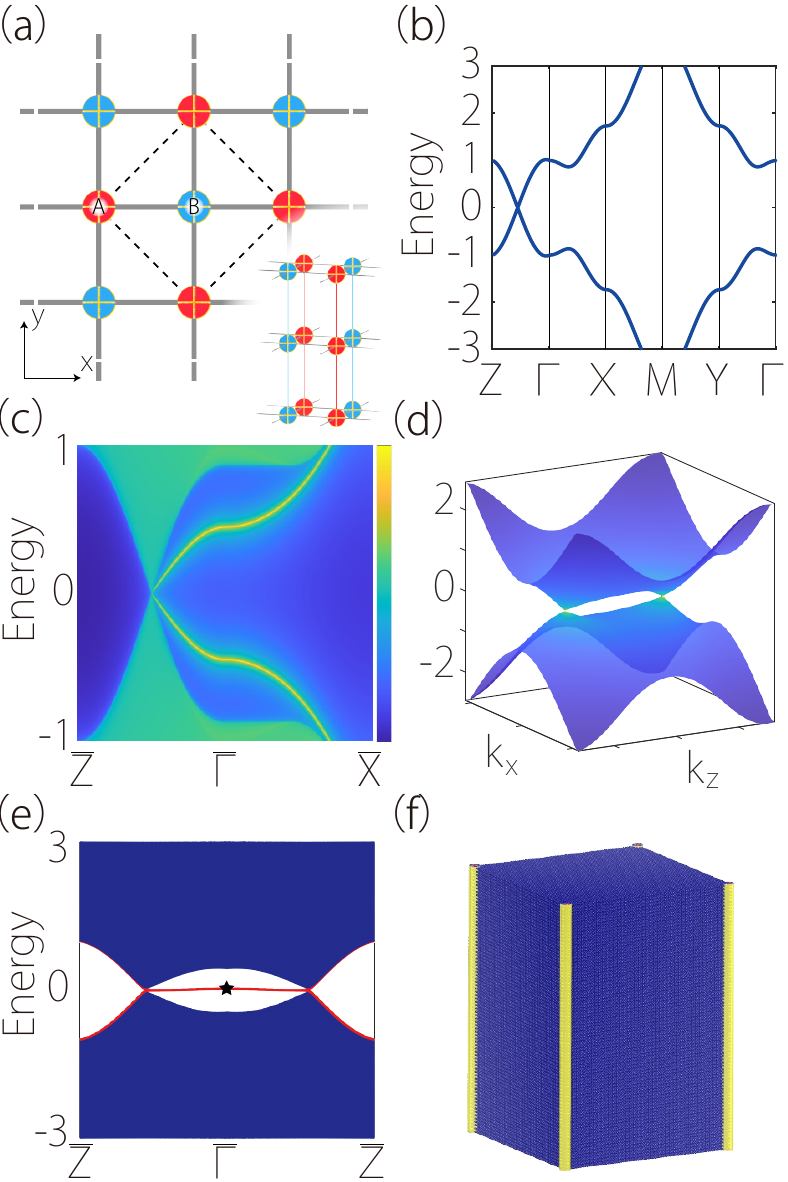}
  \caption{ (a) Illustration of the lattice model for magnetic DSM. (b) Bulk band structure. (c) Surface spectrum of the model [Eq.~(\ref{EQ:MHODSM_Htrig_4band})] along high symmetry paths for the side surface normal to $y$. (d) Surface band dispersion.
    (e) Spectrum of a 1D tube geometry shown in (f). Each side of the cross section has a length of 40 cells. The hinge modes are highlighted by the red lines.
    (f) Spatial distribution of the model marked by star in (e).
    Here, we take the parameters as $m_0 = 2, m_{1} = m_{2} = w = 0.5, v = 1, m_3 = 0.2$, and all other parameters are set to zero. }
  \label{MagDSM_4band}
\end{figure}

Since time-reversal symmetry is not a necessary condition for the existence of Dirac points, in this section we discuss hinge modes in DSMs with broken $\mathcal{T}$, i.e., in magnetic DSMs~\cite{,tang2016,hua2018}. Compared to the nonmagnetic DSMs discussed so far, magnetic DSMs exhibit  an important difference in the surface spectra. As discussed in Sec.~\ref{sec:model}, a nonmagnetic DSM has Dirac-cone type surface bands protected by the $\mathcal{T}$ symmetry. In a magnetic DSM, the $\mathcal{T}$ symmetry is broken, so the surface Dirac cone is generally gapped.

To explicitly demonstrate this point, we construct a four-band lattice model which follows the $P4/m'mm$ magnetic space group symmetry (No.~123.341 in Belov-Neronova-Smirnova notation). As shown in Fig.~\ref{MagDSM_4band}(a), we take a simple tetragonal lattice with two sites in a unit cell, labeled as A and B sites. At A site, we put two basis orbitals $|s\uparrow\rangle$ and $|s\downarrow\rangle$; and at B site, we put $|p_- \uparrow\rangle$ and 
$|p_+ \downarrow\rangle$ as basis ($p_\pm=p_x\pm ip_y$). In these four bases, the generators of the space group are represented by 
$\mathcal{PT} = -i\sigma_zs_y\mathcal{K} $, $\mathcal{M}_x = i\sigma_0s_x$, $C_{4z} = e^{i(\pi/4)(2\sigma_0-\sigma_z)s_z}$. Then, we construct the following minimal model that respects these symmetries:
\begin{equation}
  \begin{aligned}
    \mathcal{H} =
    & ~\varepsilon(\bm{k})\sigma_0s_{0} + m(\bm{k})\sigma_zs_{0} +v(\sin{k_x}\sigma_xs_z+\sin{k_y}\sigma_ys_0)\\
    & ~+w(\cos{k_x}-\cos{k_y})\sigma_xs_x,
  \end{aligned}
  \label{EQ:MHODSM_Htrig_4band}
\end{equation}
where $\varepsilon(\bm{k}) = 2\varepsilon_1\cos{k_z}+4\varepsilon_2\cos{k_x}\cos{k_y}+ \varepsilon_3\sin{k_z}$, and $m(\bm{k}) = m_0 +2m_1\cos{k_z}+4m_2\cos{k_x}\cos{k_y}+m_3\sin{k_z}$. Here, $\varepsilon_0$ and $m_0$ represent on-site energy, with $v$, $w$, $\varepsilon$'s and $m$'s being real parameters.  With properly chosen parameters, this model describes a DSM state as shown in Fig.~\ref{MagDSM_4band}(b), which has a pair of Dirac points along the $k_z$ axis. 
Here, the $\mathcal{T}$ symmetry is broken by the  $w$ term. 
If we drop the $w$ term, the model would reduce to a nonmagnetic DSM similar to the ones discussed in Sec.~II. To see this, we expand the lattice model (\ref{EQ:MHODSM_Htrig_4band}) at the $\Gamma$ point for small $k$ (the diagonal term $\sim \sigma_0s_0$ is dropped since it does not affect the topology). Then, we obtain the following $k\cdot p$ model up to $k$-quadratic terms:
\begin{equation}
  \begin{aligned}
    {H}_\text{eff}= & M(\bm{k})\sigma_zs_0+ A_zk_z\sigma_zs_0+
     Ak_x\sigma_xs_z+Ak_y\sigma_ys_0\\
                  &+B(k_x^2-k_y^2)\sigma_xs_x,
  \end{aligned}
  \label{EQ:MHODSM_k_4band}
\end{equation}
where $M({\bm k})=M_{0}-M_{1}k_{z}^{2}-M_{2}(k_x^{2}+k_y^2)$, $M_0=-m_0+2m_1+4m_2$, $M_1=m_1$, $M_2=2m_2$, $A_z=m_3$, $A=v$, and $B=w/2$.
This  model is very similar to model (\ref{EQ:DSM_k2}) except for the last term. Importantly, unlike the $k$-cubic term in (\ref{EQ:DSM_k3}), the $B(k_x^2-k_y^2)\sigma_xs_x$ term opens a gap in the 2D subsystem $H(k_x,k_y)$ for \emph{any} fixed $k_z$, including the $k_z=0$ slice, because this term derives from the $\mathcal{T}$-symmetry breaking $w$ term. It follows that the surface Dirac cone (as in Fig.~\ref{fig:DSM_k3_surf}(b-d)) will be gapped out.

This feature is confirmed by our numerical results shown in Fig.~\ref{MagDSM_4band}(c) and (d). One observes that as expected, the Dirac cone at $\bar{\Gamma}$ is removed and the surface bands are gapped. Meanwhile, the existence of the hinge modes, as corresponding to the second-order topology, is not affected. As shown in [Fig.~\ref{MagDSM_4band}(e, f)], due to the absence of the surface Dirac cone, the hinge modes can be more clearly observed in the spectrum. This could be an advantage for the detection of hinge modes.

\section{Conclusion}

In this work, we have discussed how to capture the topological boundary modes in the effective model approach to DSMs. We show that 
the $k$-cubic terms, which are often neglected in such models, are in fact essential for capturing the correct boundary-mode topology. 
Using the effective model, we can understand the evolution of surface spectrum driven by the $k$-cubic terms. Based on such understanding, we show that the surface Dirac cone and the topological hinge modes can be clearly exhibited in $\beta$-CuI. Furthermore, we show that in magnetic DSMs, the breaking of $\mathcal{T}$ symmetry can gap out the surface Dirac cone while preserving the hinge modes. This could be an advantage for the detection of hinge modes. Our finding clarifies the key features of the topological boundary modes of DSMs. It has important implications on theoretical studies on DSMs using the effective model approach. Our result also suggests $\beta$-CuI and magnetic DSMs as good candidates for probing the topological hinge modes.

%

\begin{acknowledgments}
    We thank D. L. Deng and Zhijun Wang for helpful discussions. This work is supported by the NSFC (Grants No. 12174018, No. 12074024, No. 11774018), and the Singapore Ministry of Education AcRF Tier 2 (MOE2019-T2-1-001)
\end{acknowledgments}

\appendix
   \renewcommand{\theequation}{A\arabic{equation}}
   \setcounter{equation}{0}
   \renewcommand{\thefigure}{A\arabic{figure}}
   \setcounter{figure}{0}
   \renewcommand{\thetable}{A\arabic{table}}
   \setcounter{table}{0}

  \section{Computation Method}\label{appendix:method}
  The first-principles calculations have been carried out based on the density-functional theory (DFT) as implemented in the Vienna \textit{ab initio} simulation package (VASP)~\cite{Kresse1994,Kresse1996}, using the projector augmented wave method~\cite{blochl1994} and Perdew-Burke-Ernzerhof (PBE)~\cite{perdew1996} exchange-correlation functional approach.
  The plane-wave cutoff energy was set to 500 eV.
  The Monkhorst-Pack $k$-point mesh~\cite{monkhorst1976} of size $8\times8\times 8$ was used for the BZ sampling in bulk calculations. The surface spectrum of $\beta$-CuI was calculated by constructing the maximally localized Wannier functions (MLWF)~\cite{marzari1997,souza2001} and surface Green's function methods~\cite{lopezsancho1984,lopezsancho1985} implemented in \emph{wanniertools}~\cite{wu2018a}.


  \section{Effective model with $D_{4h}$ symmetry}
  \label{appendix:AppC4}
   \renewcommand{\theequation}{B\arabic{equation}}
   \setcounter{equation}{0}
   \renewcommand{\thefigure}{B\arabic{figure}}
   \setcounter{figure}{0}
   \renewcommand{\thetable}{B\arabic{table}}
   \setcounter{table}{0}

  \begin{figure}[htb]
    \includegraphics[width=8cm]{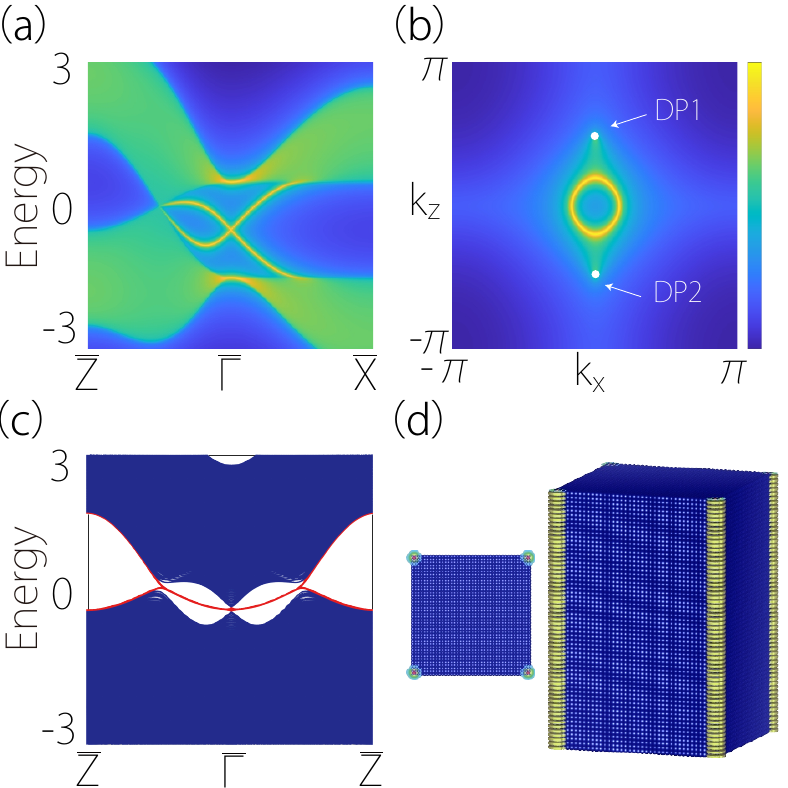}
    \caption{ (a, b) Projected spectrum and the Fermi contour on the (010) surface. The projection of bulk Dirac points are indicated by two white points. (c) Spectrum of a 1D tube geometry as shown in (d). Each side of the cross section has a width of 40 cells. The hinge modes are highlighted by the red color. (d) Spatial distribution of the hinge mode marked by star in (c).  In the calculation, we take the parameters as $C_0 = 1, C_1 = 0.25, C_2 = 0, M_0 = 1, M_1 = 0.5, M_2 = 0.5, A_0 = 1, A_1 = A_2 = 0,B_1 =B_2 = 0.5.$}
    \label{fig:DSM_C4}
  \end{figure}
  Here, we consider the effective model constrained by the $D_{4h}$ symmetry: $\mathcal{P} = \sigma_zs_0$, $\mathcal{M}_x = i\sigma_0s_x$ and $C_{4z} = e^{i(\pi/2)\hat{J_z}/\hbar} = e^{i(\pi/4)(2\sigma_0-\sigma_z)s_z}$. 
   Using the approach discussed in the main text, we find that the model expanded up to $k$-cubic order reads
  \begin{equation}
    \begin{split}
      H&=H_0+H_1,\\
      H_0&=\varepsilon(\bm{k})+M(\bm{k})\sigma_zs_0+A{(\bm{k})}k_x\sigma_xs_z-A{(\bm{k})}k_y\sigma_ys_0,\\
      H_1&=B_1k_z(k_x^2-k_y^2)\sigma_xs_x+2B_2k_xk_y\sigma_ys_x. 
    \end{split}
    \label{Eq:DSMC4}
  \end{equation}
  The functions $\varepsilon$, $M$, and $A$ have the same form as in model [Eq.~(\ref{EQ:DSM_k2})]. One can see that the main difference from model [Eq.~(\ref{EQ:DSM_k2})] is that there is one more independent parameter in $H_1$. The qualitative features of the surface and hinge spectra are the same as discussed in the main text [see Fig.~\ref{fig:DSM_C4}].

\bibliographystyle{apsrev4-1}{}

\bibliography{bibliography}


\end{document}